\documentclass[10pt,aps,prl,twocolumn,superscriptaddress,nofootinbib]{revtex4-1}
\pdfoutput=1
\usepackage{amsfonts}
\usepackage{aas_macros}
\usepackage{mathrsfs}
\usepackage{amsmath}
\usepackage{amssymb}
\usepackage{fancyhdr}
\usepackage{graphicx}
\usepackage{xspace}
\usepackage{rotating}
\usepackage[normalem]{ulem}
\usepackage{braket}
\usepackage{verbatim}
\usepackage{xcolor}
\usepackage[utf8]{inputenc}
\usepackage[medium]{titlesec}
\usepackage{bm,soul}
\usepackage[normalem]{ulem}
\usepackage{extarrows}
\usepackage{slashed}
\usepackage{isodateo}
\usepackage{graphicx}
\usepackage[dvipsnames]{xcolor}
\usepackage[bookmarksnumbered=true,bookmarksopen=true]{hyperref}
\usepackage[hmargin=.7in,vmargin=1.1in]{geometry}
\usepackage{indentfirst}
\usepackage{cancel}

\usepackage{soul}

\usepackage{amsfonts}
\usepackage{mathtools}
\usepackage{pifont}
\usepackage{mathrsfs}
\usepackage{amsmath}
\usepackage{amssymb}
\usepackage{framed}

 \usepackage{hyperref}

\newcommand{\beq}{\begin{equation}}
\newcommand{\eeq}{\end{equation}}

\newcommand{\bea}{\begin{eqnarray}}
\newcommand{\eea}{\end{eqnarray}}

\newcommand{\trd}[1]{{\textcolor{magenta}{[trd - {#1}]}}}

\begin{document}

\title{Early Formation of Supermassive Black Holes via Dark Star Gravitational Instability
}

\author{Katherine Freese}
\email{ktfreese@utexas.edu}
\affiliation{
Texas Center for Cosmology and Astroparticle Physics,
Weinberg Institute for Theoretical Physics, Department of Physics,
The University of Texas at Austin, Austin, TX 78712, USA}
\affiliation{Department of Physics, Stockholm University, Stockholm, Sweden}
\affiliation{Nordic Institute for Theoretical Physics (NORDITA), Stockholm, Sweden}

\author{George M. Fuller}
\email[]{gfuller@ucsd.edu}
\affiliation{Department of Physics, University of California, San Diego, La Jolla, CA 92093, USA}

\author{Sohan Ghodla}
\email[]{sohanghodla9@gmail.com}
\affiliation{Department of Physics and Astronomy, Colgate University, Hamilton, NY 13346, USA}

\author{Cosmin Ilie }
\email[]{cilie@colgate.edu}
\affiliation{Department of Physics and Astronomy, Colgate University, Hamilton, NY 13346, USA}

\author{Kyle S. Kehrer}
\email[]{kkehrer@ucsd.edu}
\affiliation{Department of Physics, University of California, San Diego, La Jolla, CA 92093, USA}

\author{Tanja Rindler-Daller}
\email[]{tanja.rindler-daller@univie.ac.at}
\affiliation{Department of Astrophysics, Vienna University Observatory, Vienna Int.School of Earth and Space Sciences, University of Vienna, Vienna, Austria}

\author{Evangelos I. Sfakianakis}
\email{evangelos.sfakianakis@austin.utexas.edu}
\affiliation{
Texas Center for Cosmology and Astroparticle Physics,
Weinberg Institute for Theoretical Physics, Department of Physics,
The University of Texas at Austin, Austin, TX 78712, USA}
\affiliation{Department of Physics, Harvard University, Cambridge, MA, 02131, USA}

\date{\today}

\begin{abstract}
We show that dark stars, which are dark-matter-powered stars in the early universe, can grow by accretion to masses in the range ${\cal O}\left ({10}^4\right )-{\cal O}\left ({10}^7\right)\,{M_\odot}$ 
before the general-relativistic Feynman-Chandrasekhar instability causes their dynamical collapse to black holes.
These accreting dark star configurations avoid standard stellar nuclear- and weak-interaction evolution that would lead to their demise long before they reached this supermassive size. Remarkably, this mechanism for supermassive black hole (SMBH) genesis is relatively robust to initial dark star mass, formation epoch, accretion rate and its history. 
The SMBHs produced this way can serve as seeds for even larger SMBHs $({\gtrsim}10^9\,M_\odot)$ that have been discovered at high redshift.
\end{abstract}
\maketitle

\textbf{\emph{Introduction}} -- Two 
outstanding unsolved problems in science are the nature of the dark matter 
and the origin of  
early supermassive black holes (SMBHs) with masses $\gtrsim 10^6\, M_\odot$.
SMBHs are inferred by observations to exist at epochs in the history of the universe so early that standard astrophysical origin scenarios are challenged~\citep[e.g.][]{Wang:2019,Inayoshi:2020,Lupi:2021,Bogdan:2023UHZ1,BHDorm:2024}.
Solving both of these problems may require beyond standard model (BSM) physics. 
Here we study a way these puzzles could be linked:
Dark-matter-powered massive stars, known as dark stars, which are candidates for the first stars to form when the Universe was roughly 200 million years old~\citep{Spolyar:2008dark}, and their ultimate gravitational collapse, offer an intriguing connection~\citep{Spolyar:2008dark,Freese:2010smds}.
Dark Stars are early stars
made of hydrogen and helium (primordial abundances) 
but powered by dark matter (DM).  This paper presents the first study  of general-relativistic
effects on the stability of these stars, leading to their collapse to SMBHs at high redshift.
Depending on the mass of the DM particles, we get a unique prediction for the initial mass of the resulting SMBH.

SMBHs have been observed at astonishingly high redshift.
James Webb Space Telescope (JWST)~\citep{jwstsmbh} and Chandra X-Ray Observatory~\citep{chandrasmbh} observations have been interpreted to suggest the existence of compact objects with masses ${\gtrsim}10^7\,M_\odot$ at redshift $z\sim 10$
when the universe is a scant ${\sim}480\,{\rm Myr}$ old~\citep[e.g.][]{Bogdan:2023UHZ1,BHDorm:2024,QSO1:2025arXiv250821748J,LRDs:2024ApJ...963..129M} .
Even more surprising are the massive quasars found by JWST: a striking example is UHZ-1, an enormous 
SMBH at $z\sim 10$, with a mass of about $10^{8}\,M_\odot$~\citep{Bogdan:2023UHZ1}.
There is even tentative  evidence in JWST for Black Holes (BHs) of mass $(10^4-10^5)\,M_\odot$ already at $z=25$ ~\citep{Highestz:2025A&A...701A.186M}.
Many other $z\gtrsim 6$ quasars have been observed harboring SMBHs that may be too big to have been seeded by standard stellar evolution~\citep[e.g.][]{Wang:2019,Inayoshi:2020,Lupi:2021}. 
SMBHs at $z\gtrsim6$ appear to be about three orders of magnitude more massive than what the relation between central SMBH and stellar mass in local galaxies would imply~\citep{BHdorm24}. 
Moreover, a large sub-sample of JWST's Little Red Dots (LRDs) are thought to harbor BHs that are significantly overmassive relative to their host galaxies~\citep{LRDsOvermassiveBHs:2025ApJ...985..169D}. These data may require the need for heavy BH \lq\lq seeds\rq\rq\ at high redshifts~\citep{Bogdan:2023UHZ1,ilie2023uhz1}.

 Even the most massive zero-metallicity nuclear burning stars,
 \footnote{For references on zero metallicity, a.k.a.~Pop~III, stars see, for example Refs.~\cite{Abel:2001,Barkana:2000,Bromm:2003,Yoshida:2006,OShea:2007,Yoshida:2008,Bromm:2009}.} forming at $z \sim 10-50$, could \emph{only} have seeded such gigantic SMBHs if they accreted for long periods of time at super-Eddington rates. Instead, because Pop III stars are hot and emit ample ionizing radiation, their mass accretion stops at $\sim 100 - 1000\, M_\odot$~\citep[e.g.][]{mckee2008formation}.~\footnote{In the Direct Collapse Black Hole (DCBH) scenario~\citep[e.g.][]{Loeb:1994wv,Belgman:2006,Lodato:2006hw,Natarajan:2017,barrow:2018,Whalen:2020,Inayoshi:2020} Pop~III stars can become supermassive, if they form in an atomically-cooled DM halo when: i.~a mechanism to destroy H$_2$ is invoked, e.g.~by a companion galaxy that emits abundant Lyman-Werner photons and ii.~accretion onto the star exceeds ${\sim}0.1 M_\odot/$yr.} This poses a problem for explaining early SMBHs. Even the merger of massive Pop III stars or their BH remnants is problematic, because the merging process would take too long to explain the gigantic early SMBHs inferred by the data~\citep[e.g.][]{Bogdan:2023UHZ1}. 

However, the standard picture of star formation neglects the potential effects of the large dark matter (DM) abundance in the central regions of the minihalos where the stars in the early universe are born. As the hydrogen clouds collapse, they bring in even more DM with a well-established dynamical process known as adiabatic contraction.
The DM abundance in the center becomes so large that DM particle annihilation, for example, can provide the dominant energy source for these stars~\citep{Sellwood:2005,Spolyar:2008dark}. In a later stage, there is a second source of energy due to DM from the capture of DM particles from the surroundings of the star via elastic scattering. 
 This new population of DM-powered stars have been called ``dark stars" (DSs)~\citep[][]{Spolyar:2008dark,Freese:2008ds}.
 Subsequent work includes \cite{Freese:2008cap,
 Iocco:2008,Iocco:2008cap,Gondolo:2010dmds, Freese:2010smds,
 Zackrisson:2010HighZDS, Ilie:2012,Rindler-Daller:2015SMDS, Banik:2019, Rindler-Daller:2021MNRAS, Gondolo_2022,Singh_2023,Ilie:2023JADES,ilie2025spectroscopicsupermassivedarkstar,cammelli2024}; 
 for a review see \cite{Freese:2016dark}.

 We emphasize that DSs are made almost entirely of hydrogen and helium from the early universe, with DM constituting ${\sim}0.1\%$ of the mass of the star.
 The word ``dark" refers to the stellar power source by DM, not to the make-up of the star nor its appearance. Indeed, as described further below, they can reach a luminosity of ${\sim}\,10^{10}L_\odot$ and candidates have been found in JWST data~\citep{Ilie:2023JADES,ilie2025spectroscopicsupermassivedarkstar}.    

\medskip

\noindent\textbf{\emph{Unique characteristics of dark stars enable accretion-driven growth to massive size}}--
A generic mechanism proposed in~\cite{Spolyar:2009,Freese:2010smds}
relies on a key property of DSs: their DM-derived power makes them ``puffy,''~i.e., low in density with central temperatures too low to sustain the nuclear-burning and weak-interaction evolution of a standard model star of comparable mass. This allows these DSs to grow by accretion to a very large mass and size, before they succumb to general-relativistic instability and collapse into comparably massive BHs.  In this paper, we follow for the first time the detailed evolution of these objects through that general-relativistic instability.

Any self-gravitating hydrostatic stellar configuration of mass $M$, growing in mass with accretion rate $\dot{M}$, will have that growth terminated when the evolution time-to-collapse $\tau_{\rm coll}$ at any point in its evolution falls below the mass accretion growth time scale $\tau_{\rm acc} = {M/{\dot{M}}}$. 
Usually, $\tau_{\rm coll}$ varies with mass and age. Solar-mass scale hydrogen burning stars have lifetimes of ${\sim}10$ Gyr and effectively never collapse. Standard model stars with masses in the range $8 \lesssim {M/ {M_\odot}} \lesssim 100$ have $\tau_{\rm coll} \lesssim 10\,{\rm Myr}$. The evolution of these stars is dominated by the weak interaction and neutrino emission, and the endpoint of this evolution is a Chandrasekhar-mass relativistic electron pressure-dominated core that will collapse and leave a solar-mass scale neutron star or BH remnant. Stars with masses ${\sim}100\text{--}10^4\,M_\odot$ will suffer the $e^+/e^-$ pair instability at the onset of core Carbon/Oxygen burning at roughly 2 Myr age and the star becomes unstable to collapse. 
The condition that $\tau_\mathrm{acc}<\tau_\mathrm{coll}$ implies that prodigious accretion rates would be required to grow a standard model star to ${\gtrsim}10^4\,M_\odot$. Of course, post-collapse accretion could grow the remnant BH mass.

In stark contrast to standard model stars, DSs can maintain $\tau_\mathrm{acc}<\tau_\mathrm{coll}$ throughout their accretion-driven growth, even to very large mass, and even with comparatively modest accretion rates, as we will explicitly show in this work. The key is the DM-derived energy injection inherent to DSs ~\citep{Spolyar:2008dark,Freese:2008ds}. 

 To provide a specific example in this paper, we will consider DM annihilation as the power source. However, our considerations could be qualitatively similar if we were to invoke an alternative DM power source.
 The relevant types of DM particles for powering the stars include Weakly Interacting Massive Particles (WIMPs)~\footnote{This is the case for most previous works on DSs.} and Self-Interacting Dark Matter (SIDM)~\citep{Wu:2022wzw}. 
 WIMP annihilation produces energy for the DS at a rate per unit volume 
\begin{equation} 
Q =  \langle \sigma v \rangle n_\chi^2 f m_\chi = \rho_\chi^2 f {\langle \sigma v \rangle \over m_\chi},
\label{eq:heating rate}
\end{equation}
where $m_\chi$ is the DM particle mass, $\rho_\chi$ is the DM mass-energy density, and $f\sim 2/3$ is the fraction of the DM mass deposited as heat into the star. Per mass of the reactants, DM annihilation provides ${\sim}70\%$ efficiency in energy injection (the rest is lost to neutrinos), while hydrogen burning in normal stars gives only ${\sim}1 \%$. We will consider a variety of DM particle masses $1 \, {\rm GeV}\le m_\chi  \le 10^5$ GeV \footnote{Unitarity bounds imply the WIMP mass is $m_\chi\lesssim 100$~TeV~\cite{Griest:1990}.} and use the standard thermally-averaged annihilation cross section $\langle \sigma v \rangle = 3 \times 10^{-26}\,{\rm cm}^3/{\rm s}$, unless noted otherwise. A weak interaction annihilation cross section in the  early universe of roughly this value gives rise to a significant DM relic abundance today. 
  Since the heating rate scales as ${\langle \sigma v \rangle / m_\chi}$,  varying the WIMP mass in the simulations is equivalent to fixing the mass and instead
 varying the cross section. 
 Thus, DSs can exist with WIMP masses and annihilation cross sections spanning several orders of magnitude in these quantities.

Starting from their inception at ${\sim}1 M_\odot$, DSs accrete mass from their surroundings to become supermassive stars (SMDSs), 
some reaching masses of ${\gtrsim}10^6 M_\odot$ and luminosities ${\gtrsim}10^{9} L_\odot$, depending on the halo environment~\cite{Freese:2010smds, Rindler-Daller:2015SMDS}. 
In fact, SMDSs are giant diffuse objects, with radii ${\sim}10\mbox{--}30$ a.u. and central densities ${\sim}10^{-4}$ g/cm$^3$.
DSs remain cool throughout their evolution, with maximum $T_\text{eff} \sim (2\mbox{--}3)\times 10^4$ K, implying there is no ionizing radiation feedback to cutoff accretion-driven growth. By contrast, a standard Pop III star with the same mass as a DS will have a much higher surface temperature and concomitant radiation pressure that could suppress accretion~\citep[e.g.][]{mckee2008formation}.

Dark stars continue to grow in mass as long as there is DM fuel. If at some point, a SMDS leaves the DM-rich region, e.g. when the original minihalo hosting it merges with others, the energy injection from DM annihilation  ceases and  then  the SMDS 
becomes unstable to collapse to a SMBH~\citep[see][]{Spolyar:2008dark,Freese:2008cap,ilie2023uhz1}. 
\medskip

\textbf{\emph{General relativistic instability}} -- Accretion-driven mass growth cannot go on indefinitely. As the DS mass grows, the fraction of pressure support stemming from the radiation field increases. The run of density, temperature and pressure in these configurations will be closely that of an index $n=3$ polytrope. This is borne out with calculations of DS structure performed in \cite{Rindler-Daller:2015SMDS, Rindler-Daller:2021MNRAS}, using the 1D stellar evolution code \texttt{MESA}~\citep{Paxton2011,MESA2025}. The temperature gradient in these DSs will be close to the adiabatic gradient. They will be fully convective with roughly constant entropy-per-baryon in units of Boltzmann's constant $k_\mathrm{b}$, $s\sim 1000(M/10^6\,M_\odot)^{1/2}$. These stars are completely Newtonian in their structure, with total energy, gravitational energy plus thermal internal energy, close to zero. As such they are ``trembling on the verge of instability''\footnote{Phrase attributed to W.A.\ Fowler.} and  will eventually suffer the general-relativistic (GR) Feynman-Chandrasekhar instability~\cite{fh1962,iben1963,fowler1964,fww86,fc,Feynman:1996kb}.

\medskip

Though the radiation-dominated DS is Newtonian in its structure, its stability is determined by the first non-linear correction of GR. Instability ensues when the pressure-averaged value of the adiabatic index falls below
\begin{equation} \label{eq:schwarzschild}
    \left<\Gamma_1\right><\frac{4}{3}+{\cal{O}}\left(\frac{r_s}{R}\right),
\end{equation}
where $R$ is the stellar radius with corresponding Schwarzschild radius $r_s$, and where the adiabatic index is $\Gamma_1=(\partial\ln P/\partial\ln\rho)|_s$, with $P$ and $\rho$ being the total pressure and density, respectively. In the radiation-dominated conditions of a supermassive star, $\Gamma_1\approx 4/3 + \beta/6$, where $\beta\ll1$ is the ratio of non-relativistic gas pressure to the total pressure. 

For supermassive stars, such as SMDSs, instability and collapse will occur when the central density reaches 
\begin{equation}
    \rho_{\mathrm{c}, \text {crit}} \approx 3.98\left(\frac{0.59}{\mu}\right)^3\left(\frac{10^5 M_{\odot}}{M}\right)^{7 / 2} \mathrm{g\,cm}^{-3},
    \label{eq:critCentralDensity}
\end{equation}
where $\mu\approx 0.59$ is the mean molecular weight (atomic mass units per particle) for a fully ionized primordial abundance gas and $M$ is the total stellar mass~\citep[e.g.][]{Kehrer:2024dga}. Note that nuclear burning would inevitably lead to larger nuclei and that would raise $\mu$ and so lower $\rho_{\mathrm{c}, \text{crit}}$. At a given temperature, a larger $\mu$ also means a lower non-relativistic gas contribution to the total pressure, i.e., a lower $\beta$. Both effects drive the star closer to instability. Note that these effects would be absent for a SMDS that experiences {\it no} nuclear burning prior to collapse.

With primordial composition, in either a purely standard model supermassive star or a SMDS, the injection of energy from nuclear burning {\it after} the instability will be too slow to halt the collapse to a black hole \cite{fww86}. This, plus the prodigious neutrino emission during the collapse, will guarantee that a substantial portion of the supermassive progenitor will become a black hole \cite{shifuller, 1998ApJ...502L...5F, 1997ApJ...487L..25F}. It is worth noting that the presence of DM in supermassive stars leads to a change in Eq.~\eqref{eq:critCentralDensity}, but this is only significant for a DM mass fraction considerably larger than what is envisioned for these DSs \cite{Kehrer:2024dga}; DSs only have ${\lesssim}0.1\%$ of their mass in DM, as mentioned earlier.

 \medskip

\textbf{\emph{Simulation Results}} -- 
We simulate the evolution of DSs using \texttt{MESA}, as described in Refs.~\cite{Rindler-Daller:2015SMDS, Rindler-Daller:2021MNRAS}. 
We run models with initial DS masses in the range $1-10~M_\odot$ and find that the downstream evolution of the DS, including its mass at instability, is not much affected by the choice of initial mass. 
Our simulations assume spherical mass accretion\footnote{Since $\dot{M} \propto T^{3/2}$, ambient accretion rates in early halos with primordial gas ($T \sim 200\mbox{--}300$ K) are more than two orders of magnitude higher than in present-day star-forming regions ($T \sim 10$ K); see e.g.~\cite{Bromm:2003}.}  at the following constant rates: $\dot M =\{ 10^{-3}, 10^{-2},10^{-1} \} \,  M_\odot$ yr$^{-1}$. 

As a DS accretes mass over time, it remains in thermal and hydrostatic quasi-equilibrium. \texttt{MESA} simulations show that at large enough mass the DS pressure support will be dominated by radiation and the run of density and pressure will be roughly that of an index $n=3$ polytrope. The evolution of the DS luminosity and surface temperature is such that the DS shines at roughly the Eddington limit as soon as $M_{DS}\gtrsim10^3M_\odot$, while $T_\text{eff}\lesssim50,000$~K, even for the most massive SMDSs~\citep{Rindler-Daller:2015SMDS}.

\begin{figure}
\centering
\includegraphics[scale=0.85]{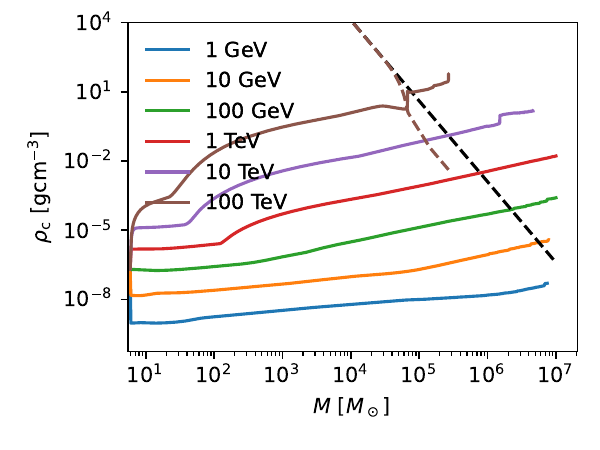}
\caption{Evolutionary tracks (solid curves) for DSs as calculated with \texttt{MESA} are shown on a plot of the central density $\rho_{\rm c}$ {\it vs.} the stellar mass $M$. 
These calculations assume mass accretion 
at a constant rate of $\dot M = 10^{-2} M_\odot$ yr$^{-1}$. Different colors correspond to different DM particle masses $m_\chi  = 1,10,100,10^3,10^4,10^5$ GeV (bottom to top). The dashed curves show the critical density for the onset of GR instability and collapse, Eq.~\eqref{eq:critCentralDensity}. Here, the mean molecular weight $\mu$ is computed self-consistently as a function of $M$ for each star and deviates from the primordial value $\mu=0.59$ (black-dashed curve) only in the case of $m_\chi=100$ TeV (brown-dashed curve). 
The intersection of the solid and dashed lines for a given color marks the GR instability point in the DS's evolution.} 

\label{fig:rhocrit_vs_M}
\end{figure}

Fig.~\ref{fig:rhocrit_vs_M} shows the evolution of the central baryonic density of DSs powered by the annihilation of DM particles with the indicated rest masses. Here we neglect the possible role of DM capture, and only focus on DSs where the DM reservoir is provided by adiabatic contraction. All of these evolutionary tracks are calculated with a constant accretion rate of $\dot M = 10^{-2} M_\odot$ yr$^{-1}$. 
The DS becomes unstable when its central density becomes equal to the critical density given by Eq.~\eqref{eq:critCentralDensity} for a given value of its mass. 
The dashed lines in Fig.~\ref{fig:rhocrit_vs_M} show this critical density. 
The intersection of the solid and dashed lines gives the DS mass at which collapse to a BH ensues, i.e.~the critical mass ($M_\text{crit}$), as explained below.

 The dashed line corresponding to the critical central density in Eq.~\eqref{eq:critCentralDensity} is the same for all DM masses except the highest, $m_\chi=100$ TeV.
For lower stellar and DM particle masses, the star is only powered via DM annihilation and therefore $\mu$ is constant at its initial primordial value $0.59$. From Eq.~\eqref{eq:heating rate}, one can see that for the same $\rho_\chi$, a variable $m_\chi$ can impact the efficiency of DM heating. In particular, a DM halo made of heavier DM particles (the case of $m_\chi = 100$ TeV) will supply less energy to the star, causing it to contract and heat up.
This leads to an initiation of hydrogen burning, in turn causing $\mu$ to rise 
to $\mu\simeq 1.8$ (shown by the brown-dashed line, $m_\chi=100$ TeV, in Fig.~1).  Note that the black-dashed curve is identical to (and hidden under) the brown-dashed curve in Fig.~1 at the highest central densities.

\begin{figure}
\includegraphics[scale=0.85]{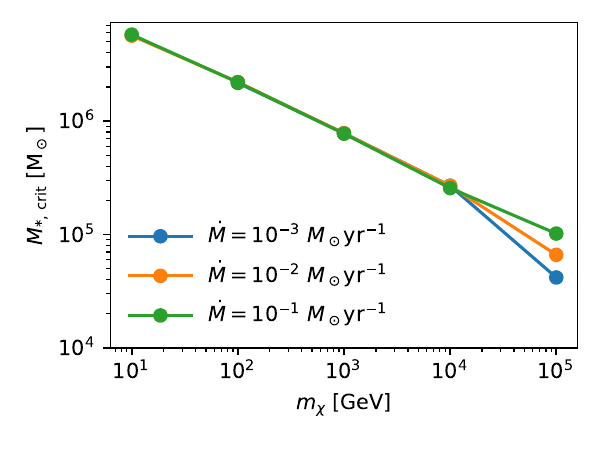}
\caption{The critical mass of a DS at the onset of collapse as a function of the DM particle mass for various mass accretion rates. 
The critical mass of the DS can be used as an estimate for the mass of the resulting BH.
We see that the critical mass is independent of the accretion rate, except for the case of $m_\chi=100$ TeV, as explained in the text.
}\label{fig:MDM_vs_MCrit}
\end{figure}

In Fig.~\ref{fig:MDM_vs_MCrit} we plot the critical mass of a DS as a function of the DM mass for various mass accretion rates $\dot M$. 
We only notice a difference in outcome for a DM mass larger than 10 TeV. 
 For $m_\chi\gtrsim 10\,{\rm {TeV}} $, the core begins hydrogen burning, as discussed above. By the time the central density reaches the instability point, $\mu$ will rise to a value larger than that of primordial gas. However, depending on the accretion rate, the star will gain mass during this time. Therefore, the final value of the critical mass rises with a rising value of $\dot M$. The DS mass at which the GR instability appears $M_{*,\mathrm{crit}}$ can be fit by a simple power-law, for most of the ($m_X,\langle \sigma v\rangle$) WIMP parameter space:
\begin{equation}
M_{*, \rm crit}(m_\chi)\simeq 1.7 \times 10^{7} \,\left ( {{m_\chi 
\over {\rm GeV}}}  {3 \times 10^{-26}\,{\rm cm}^3/{\rm s}\over \langle \sigma v \rangle}\right )^{-0.45} M_\odot \, .
\label{eq:Mcritvsmchi}
\end{equation}
As expected from Eq.~\ref{eq:heating rate}, the ratio $m_\chi/\langle\sigma v\rangle$ will determine all the properties of DSs, including the mass at which they collapse.  In what follows we will restrict our attention to $\langle \sigma v \rangle = 3 \times 10^{-26}\,{\rm cm}^3/{\rm s}$. In this case, for $m_\chi\gtrsim 10$ TeV the above fit fails. In particular, for $m_\chi=100$ TeV the critical mass becomes $M_{*, \rm crit}(m_\chi) / M_\odot \simeq 4\times 10^4, 7\times 10^4, 10^5$ for $\dot M/ M_\odot ,\mathrm{yr}^{-1} = 10^{-3}, 10^{-2}, 10^{-1}$, respectively.
We do not pursue larger DM masses, as $m_\chi=100$ TeV is already close to the unitarity bound of WIMPs~\cite{Griest:1989wd}.

\medskip

\textbf{\emph{Conclusion and Outlook}} -- 
In this \textit{Letter} we have studied for the first time the self consistent evolution of accreting DSs along with general relativistic stability considerations that inevitably lead to collapse of these objects to SMBHs. We find that the key feature of accreting DSs that enable them to grow to supermassive size prior to collapse is that they are powered by dark matter (DM) and therefore \lq\lq puffy\rq\rq\ with relatively low density and temperature compared to a normal star of comparable mass. This enables DSs to avoid standard model nuclear burning and weak interaction evolution and associated neutrino losses that bring down accreting normal stars before they can reach supermassive size.
Our main result is encapsulated by Eq.~\eqref{eq:Mcritvsmchi}. Specifically, for DM mass of $(10 , 100, 1000, 10^4)$ GeV, the mass of the DS that will collapse to a SMBH is given by $\approx(6\times 10^6, 2\times 10^6, 7.6 \times 10^5 ,2.6\times 10^5) M_\odot$.
For a given DM particle mass, the resulting initial mass of the SMBH is  robust, independent of initial DS mass and formation epoch, accretion rate and history. 
It is interesting to note that for a given DM mass, the resulting SMBHs from the collapse of DSs will have a unique  mass.
An observational imprint of this result could then be used to tell us about the DM particle mass. 
As a caveat, DSs of lower mass may already run out of DM fuel before ever reaching GR instability, producing lighter BHs. In any case, for a given DM mass, we have found the largest SMBH that results from DSs. These SMBHs can then serve as seeds for the even larger SMBHs ${\gtrsim}10^9 M_\odot$ that have been discovered at high redshift, including the quasars found by JWST.

Further, we note that the key ingredient that differentiates DSs from standard model stars is that they are powered by a different heat source altogether -- 
something from beyond the standard model of particle physics.
Specifically, we have considered DM annihilation as the power source.  However, other BSM physics, including other DM candidates, might provide an energy source for DSs that might produce DS evolution that might be broadly similar to that found here.  

Remarkably, since DSs were originally proposed in 2007, they continue to provide solutions to new astronomical puzzles as they are discovered.  In particular, one conundrum posed by JWST is the enormous stellar mass required to explain the very bright ultra-compact ($r_e\lesssim 400$pc) objects seen at high redshift, if they are interpreted as early galaxies~\citep[see][and references therein]{JWSTBLueMonsters:2025A&A...694A.286F,Ziegler:2025plz}.
However, one SMDS can be as bright as an entire early galaxy of stars, so it stands to reason that some of these inexplicably bright early objects may instead be DSs. Indeed, predicted spectra
and morphologies of SMDSs are a good match to both photometric and spectroscopic data for some of these JWST objects~\citep{Ilie:2023JADES,ilie2025spectroscopicsupermassivedarkstar}. Future work will study the gravitational wave (GW) production during the collapse of a SMDS to a SMBH. Previous work has shown that for standard model supermassive stars, even a small amount of initial anisotropy in the collapsing matter distribution can emit prodigious GW radiation~\citep{lifuller}, making the equivalent process in DSs another promising signal for them.
Dark Stars can explain the gigantic early luminous objects seen by JWST, they can also explain the puzzling existence of Supermassive Black Holes, and their discovery can teach us about the nature of dark matter.

\medskip

\noindent\textbf{Acknowledgments} -- 
K.F. and E.I.S. are grateful for support from
the Jeff \& Gail Kodosky Endowed Chair in Physics at the University of Texas. 
K.F.  acknowledges
support from the Swedish Research Council (Contract
No. 638-2013-8993). C.I. acknowledges funding from Colgate University via the Research Council (Grant No. 821028) and the Picker Interdisciplinary Science Institute (Grant No. 826837). GMF and KK acknowledge support from National Science Foundation (NSF) Grants  No.\ PHY-2209578 and No.\ PHY-2515110 at UCSD and the {\it Network for Neutrinos, Nuclear Astrophysics, and Symmetries} (N3AS) NSF Physics Frontier Center, NSF Grant No.\ PHY-2020275, and the Heising-Simons Foundation (2017-228). T.R.-D. acknowledges the support by the Austrian Science Fund FWF Grant No. \ P36331-N.
 We acknowledge the use of Colgate’s Turing Supercomputer 
(Partially supported by NSF grant OAC-2346664).

\bibliographystyle{apsrev4-1}
\bibliography{DSGR}

\end{document}